\documentclass[conference,10pt,USenglish]{IEEEtran}

\usepackage{babel}
\usepackage[babel]{microtype}

\usepackage{amsmath}
\usepackage{amssymb}
\usepackage{amsthm}
\usepackage{mleftright}
\usepackage[keeplastbox]{flushend}
\usepackage{algorithm}
\usepackage{algpseudocode}
\usepackage{dsfont}
\usepackage{siunitx}

\usepackage[backend=biber,style=ieee,url=false,doi=false,isbn=false,eprint=false,maxnames=3]{biblatex}
\usepackage[babel]{csquotes}

\theoremstyle{plain}%
\newtheorem{theorem}{Theorem}%
\newtheorem{lemma}[theorem]{Lemma}

\theoremstyle{remark}

\theoremstyle{definition}

\usepackage{float}
\usepackage{graphicx}
\usepackage{booktabs}

\usepackage{xcolor}

\usepackage{tikz}
\usetikzlibrary{shapes.geometric, arrows, plotmarks, arrows.meta}
\usepackage{pgfplots}

\pgfplotsset{compat=newest}

\newlength\figureheight
\newlength\figurewidth

\bibliography{literature.bib}
\renewcommand{\bibfont}{\footnotesize}

\DeclareMathOperator{\tr}{tr}
\DeclareMathOperator{\diag}{diag}
\newcommand{\mat}{\boldsymbol}
\renewcommand{\mathbf}{\boldsymbol}
\newcommand{\trace}[1]{\tr(#1)}

\newcommand{\norm}[1]{\ensuremath{\left\|#1\right\|}}
\newcommand{\complexes}{\ensuremath{\mathds{C}}}

\graphicspath{ {images/} }

\begin{document}
\title{Achievable Physical-Layer Secrecy in Multi-Mode Fiber Channels using Artificial Noise}

\author{\IEEEauthorblockN{Eduard~Jorswieck, Andrew~Lonnstrom, Karl-Ludwig~Besser}
\IEEEauthorblockA{Institute for Communications Technology\\ TU Braunschweig, Germany\\ (e.jorswieck, a.lonnstrom, k.besser)@tu-bs.de}
\and 
\IEEEauthorblockN{Stefan~Rothe,~Juergen~W.~Czarske}
\IEEEauthorblockA{Chair of Measurement \\and Sensor System Techniques, \\ TU Dresden, Germany \\ (stefan.rothe, juergen.czarske)@tu-dresden.de}}

\maketitle

\begin{abstract}
Reliable and secure communication is an important aspect of modern fiber optic communication. In this work we consider a multi-mode fiber (MMF) channel wiretapped by an eavesdropper.
We assume the transmitter knows the legitimate channel, but statistical  knowledge of the eavesdropper’s channel only.
We propose a transmission scheme with artificial noise (AN) for such a channel.
In particular, we formulate the corresponding optimization problem which aims to maximize the average secrecy rate and develop an algorithm to solve it. We apply this algorithm to actual measured MMF channels. As real fiber measurements show, for a 55 mode MMF we can achieve positive average secrecy rates with the proper use of AN. Furthermore, the gain compared to standard precoding and power allocation schemes is illustrated. 
\end{abstract}


\section{Introduction}

Single mode fiber optical transmission systems are approaching the limit of their theoretical communication limits \cite{Arik2014} and multi-mode fibers (MMF) are proposed. They are well studied from theoretical as well as in practical aspects \cite{Mizuno2016, Ryf2015, Mizuno2014}. Similar to multiple antenna wireless systems, MMF exploits the spatial domain in order to improve the achievable data rates. Therefore, multiple-input multiple-output (MIMO) signal processing techniques can be used to separate the individual modes \cite{Arik2014}. Thereby, precoding as well as receiver methods can be applied. These comprise linear and non-linear precoding at the transmitter side \cite{Koike2017,Abouseif2020}. 

The reliable transmission of high data rate traffic is only one important key performance metric. Another important design dimension is to keep the data confidential from eavesdropping attacks. Typically, data is secured at the higher communication layers either through symmetric key or asymmetric key cryptography. Asymmetric key cryptography is shown to be weak against quantum computers which could "efficiently" solve the prime-factorization problem which forms the basis for many public key algorithms \cite{Gidney2021}. Classical information theoretic coding schemes can provide post-quantum security. Already in \cite{Wyner1975} it was shown that a proper designed wiretap-code can be designed to achieve reliable and confidential data transmission at the same time. Much research has been done in the area of physical layer security since then. In \cite{Khisti2010}, the multiple-input multiple-output multiple eavesdropper (MIMOME) scenario was investigated and secrecy capacities were derived for wireless antenna systems. In \cite{Winzer2011}, the MIMO capacities were examined for MMF and in \cite{guan2015} the secrecy capacities were examined for mode dependent loss limited MMF systems.
Since the action of the wiretapper is not known to the transmitter, it does not have channel state information (CSI) about the wiretapper channel. It is well known that with imperfect CSI at transmit side, artificial noise (AN) is required to improve the achievable secrecy rates \cite{Negi2005}. 

In \cite{Rothe2020}, a model for wiretapping attacks in MMF was developed which is experimentally verified. The proposed scheme applies inverse precoding and exploits the fact that the channel matrices from the transmitter (Alice) to the legitimate receiver (Bob) and to the wiretapper (Eve) differ significantly \cite[Figure 1]{Rothe2020}. In particular the so called mode-dependent losses (MDL) on the diagonal of the channel matrices vary. Even if Eve is able to overhear the calibration phase signals, she cannot obtain any useful information about the messages. The precoding is realized by a spatial light modulator (SLM). The resulting SNR advantage reported in \cite{Rothe2019a} was further exploited by adding white Gaussian noise to the sent message after calibration. 

In this paper, we develop techniques for enhancing the achievable secrecy rate in a MMF through a systematic use of AN. In contrast to recent state of the art \cite{guan2015} where both the legitimate channel and the eavesdropper channel are modeled by statistically independent random unitary matrices which additionally suffers from MDL, our study is based on a real channel matrix measurement for the channel to Bob, and a trace preserving variation of this channel to Eve. This model is more realistic since Eve has the same total receive power as Bob. The advantage is based on proper chosen modes from the channel to Bob while weak modes are filled with AN. We formulate the mode assignment, signal and AN power allocation as an optimization problem. Since the transmitter does not know the exact channel matrix to Eve, it maximizes the average secrecy rate. We propose an algorithm to solve the problem approximately and show the performance gain compared to the baseline. Furthermore, we compare the achieved secrecy rates with lower and upper bounds. 

\subsection{Notation}

Throughout this work, we use bold capital letters (e.g. $\mathbf{H}$) to refer to matrices of dimension $m \times n$. Bold lower case letters (e.g. $\mathbf{x}$) are used to refer to column vectors of dimension $n \times 1$. The expected value is denoted $\mathbb{E}[\cdot]$. Complex circularly symmetric Gaussian random variables with mean $\mathbf{\mu}$ and covariance matrix $\mathbf{\Sigma}$ are denoted as $\mathcal{CN} (\mathbf{\mu}, \mathbf{\Sigma}) $. The trace and determinant of a matrix $\mathbf{X}$ is written as $\tr(\mathbf{X})$ and $ \lvert \mathbf{X} \lvert$, respectively. $\mathbf{X}^T$ and $\mathbf{X}^\dag$ are the transpose and Hermitian transpose of $\mathbf{X}$, respectively. A square diagonal matrix with elements of $\mathbf{x}$ is represented as $\diag(\mathbf{x})$ where all off-diagonal elements are 0. Finally, $ \left[ x \right] ^+$ is the maximum of $x$ and 0.

\section{System Model and Problem Statement}

We consider a complex-valued MMF wiretapping system where $N$ represents the number of modes in the fiber, where $\mat{x}\in\complexes^{N}$ is signal vector sent by Alice \cite{Lonnstrom2017}. The received signals at Bob $\mat{y}\in\complexes^{N}$ and Eve $\mat{z}\in\complexes^{N}$ are modelled as
\begin{equation}
    \mat{y} = \mat{H} \mat{x} + \mat{n}_b, \qquad \mat{z} = \mat{G} \mat{x} + \mat{n}_e, \label{eq:sigmod}
\end{equation}
with complex MMF $N \times N$ channel matrices $\mat{H}$ and $\mat{G}$, with complex statistical independent proper Gaussian distributed additive noise $\mat{n}_b$ and $\mat{n}_e$ zero-mean with variance $\sigma^2$. We impose a power constraint on the transmit signal via the transmit covariance matrix $\mat{Q} = \mathbb{E}[\mat{x} \mat{x}^H]$ as $\trace{\mat{Q}} \leq P$ which translates to the average light intensity sent into the MMF. Similar to \cite{Lonnstrom2017}, we assume that Alice and Bob have perfect CSI about the channel to Bob, obtained from the training phase reported in \cite{Rothe2020}, while they only have statistical CSI about the channel to Eve. The wiretapper Eve has acces to perfect CSI of both channels. 

The channel from Alice to Bob is based on a measured channel and fixed to $\mat{H}$, while we model Eve's channel similar to \cite{guan2015} where the effect of coupling the light out of the fiber (e.g. through bending) is modeled in a diagonal MDL matrix. This per mode attenuation/loss is given as $\mathbf{L}^{\frac{1}{2}} = \diag(l_1, ... , l_{N})$, where $l_i$ are the losses at each mode. The MDL is characterized by the ratio between the lowest and highest mode attenuations (i.e. $\text{MDL}_{dB} = 10 \log_{10} (\frac{l_{\text{max}}}{l_{\text{min}}})$). The remaining elements of the loss matrix are randomly distributed between $l_{\text{max}}$ and $l_{\text{min}}$ as described in \cite[Sec. II-B]{guan2015}. 

The effective channel to Eve is then modelled as 
\begin{equation}
\mathbf{G} = \mat{L}^{\frac{1}{2}} \mat{H U}_e,  \label{eq:chanEve}
\end{equation} 
where $\mathbf{U}_e$ is a random unitary matrix. Note that the channel from Alice to Bob $\mathbf{H}$ occurs because we do not have direct measurements for an eavesdropper's channel. We therefore apply a random unitary rotation $\mathbf{U}_e$ to each realization of Eve's channel.  

We define the SVD of $\mathbf{H}$ and $\mathbf{G}$ as
\begin{align}
\text{svd}(\mathbf{H}) &= \mathbf{W} \mathbf{D T}^\dag, \label{eq:svdBob} \\
\text{svd}(\mathbf{G}) &= \mathbf{U \Sigma V}^\dag, \label{eq:svdEve}
\end{align}
where $\mathbf{D}$ is a diagonal matrix containing the singular values, $[d_1, \cdots, d_N]$ (with $d_1 \geq \cdots \geq d_N $), of the channel from Alice to Bob and $\mathbf{\Sigma}$ is a diagonal matrix containing the singular values, $[\sigma_1, \cdots, \sigma_N]$ (with $\sigma_q \geq \cdots \geq \sigma_N$), of the channel from Alice to Eve. 

Without perfect CSI, the optimal transmit strategy is to apply AN in addition to the signal. Denote the signal containing the useful message by $s$ obtained from a Gaussian codebook (with zero-mean and variance one) and the AN signal by $\alpha$ also Gaussian distributed with zero-mean and variance one). Convert the scalar signal $s$ in $S$ parallel signals $\mat{s}\in\complexes^{S}$ as well as the scalar AN $\alpha$ in $A$ parallel signals $\mat{\alpha}\in\complexes^{A}$. The transmitter generates the transmit vector $\mat{x}$ as follows 
\begin{eqnarray}
    \mat{x} = \mat{F} \mat{s} + \mat{E} \mat{\alpha}   \label{eq:precod}
\end{eqnarray}
where we can easily compute the transmit covariance matrix 
\begin{eqnarray}
    \mat{Q} = \mathbb{E} [\mat{x} \mat{x}^H] = \mat{F} \mat{F}^H + \mat{E} \mat{E}^H \label{eq:Q}. 
\end{eqnarray}
Note that the dimensions of the signal $S$ and the AN $A$ are flexible. 
We denote the transmit covariance of the signal by $\mat{Q_s}= \mat{F} \mat{F}^H$ and of the AN by $\mat{Q_a} = \mat{E} \mat{E}^H$. 
For a given set of transmit covariance matrices for the signal and artificial noise, $\mat{Q_s}$ and $\mat{Q_a}$ respectively, we calculate the secrecy rates based on 
\begin{align}
R_b &= \log \lvert \mat{I} + [ \sigma^2 \mat{I} + \mat{H} \mat{Q_a} \mat{H}^\dag]^{-1}   [ \mat{H} \mat{Q_s} \mat{H}^\dag] \rvert, \label{eq:capacityBob} \\
R_e &= \log \lvert \mat{I} + [\sigma^2  \mat{I} + \mat{G} \mat{Q_a} \mat{G}^\dag]^{-1}   [ \mat{G} \mat{Q_s} \mat{G}^\dag] \rvert, \label{eq:capacityEve}
\end{align}
and secrecy rate is then calculated as \cite{Bloch2011}
\begin{align}
R_s = \left[ R_b - R_e \right] ^+ . \label{eq:secRateObj}
\end{align}
In order not to disturb the decoding at Bob, the AN is restricted to the null-space of the useful signal, i.e., 
\begin{eqnarray}
    \mat{F} \mat{E}^H = \mat{0}  \label{eq:prop}.
\end{eqnarray}
Furthermore, we apply SVD precoding for the known channel from Alice to Bob $\mat{H}$. 
The secrecy rate - as modelled and estimated at Alice - can be written as
\begin{multline}
\hat{R}_s = \bigg[ \sum_{i=1}^k \log(1 + d_i^2 p_i) \\ 
- \mathbb{E} \left[ \log \lvert \mathbf{I + U \Sigma V^\dag[Q_s + Q_a] V \Sigma^\dag U^\dag} \rvert \right]\\
+  \mathbb{E} \left[ \log \lvert \mathbf{I + U \Sigma V^\dag Q_a V \Sigma^\dag U^\dag} \rvert \right] \bigg]^+ \label{eq:objSecRate}
\end{multline}
This approach results in the following programming problem: The objective is to choose the power allocation distribution for $\mat{Q_s}$ and transmit covariance matrix $\mat{Q_a}$ such that \eqref{eq:secRateObj} is maximized, i.e. 
\begin{align}
\begin{split}
\max_{\mat{Q_s > 0},  \mat{ Q_a \geq 0}}  &\hat{R}_s \\ 
\text{s.t.} \quad&\tr(\mat{Q_s + Q_a}) \leq P_{t}
\end{split}
\end{align}

\section{Precoding Schemes and Bounds}

In this section, a number of different precodings are reviewed. Furthermore, we derive lower and upper bounds which serve for performance comparisons in the later numerical illustrations. 

\subsection{Waterfilling Algorithm}

One of the popular baseline schemes is the waterfilling algorithm which is optimal for a peaceful system with perfect CSI at transmitter and receiver. It solves the sum rate maximization (only $R_b$ in (\ref{eq:secRateObj})).

\subsection{Secrecy Rate Bounds}

Since we do not have a closed form solution to the optimization problem we will then look at the upper and lower bounds on the secrecy rate to remove the $\mathbb{E}[\cdot]$ w.r.t $\mathbf{U}$ in \eqref{eq:objSecRate}. Our uncertainty is related to the rate achieved by Eve $R_e$. Therefore, the next lemma provides upper and lower bounds.

\begin{lemma}
The maximum rate for Eve is bounded by
\begin{equation}
R_e  \leq   \sum_{i}^{N} \log (\alpha_i + p_{N+1-i}) - \sum_{i}^{N} \log (\alpha_i + \beta_{i}) 
\end{equation}
where $p_i$ and $\beta_i$ are the ordered eigenvalues of $\mathbf{Q_s}$ and $\mathbf{Q_a}$ respectively, and $\alpha_i$ are the ordered inverse eigenvalues of $\mathbf{G}$. The lower bound reads as
\begin{equation}
    R_e \geq \sum_{i}^{N} \log (\alpha_i + p_{i}) - \sum_{i}^{N} \log (\alpha_i + \beta_{i}) .
\end{equation}
\end{lemma}
The proof can be found in \cite{Lonnstrom2021}.

\subsection{Ergodic Secrecy Rate}

If Alice had perfect CSIT for both $\mat{H}$ and $\mat{G}$, then we could solve for the optimal covariance matrix $\mat{Q}$ to maximize the difference between the channel capacities of  Bob and Eve. According to \cite{Bustin2009} solving the following optimization problem will result in an upper bound for the secrecy rate with perfect CSIT for both $\mat{H}$ and $\mat{G}$. Since perfect CSIT for Eve's channel is not available, a comparison could be done with a lower bound of the ergodic secrecy rate of the channel, i.e.
\begin{align}
R_{erg} & = \log \lvert \mathbf{ I +  Q H^\dag H} \rvert - \mathbb{E}[\log \lvert \mathbf{ I +  Q G^\dag G} \rvert] \label{eq:ergSecRate} \\
 & \geq \log \lvert \mathbf{ I +  Q H^\dag H} \rvert - \log \lvert \mathbf{ I +  Q \mathbb{E}[G^\dag G]} \rvert \label{eq:jenSecRate}
\end{align}
where the step from (\ref{eq:ergSecRate}) to (\ref{eq:jenSecRate}) is done using Jensen's inequality. Solving for $\mathbf{Q}$, based on \cite{Nguyen2020}, in (\ref{eq:jenSecRate}) will result in a more \enquote{pessimistic} ergodic secrecy rate. 

\section{Proposed AN Precoding Algorithm}

The basic idea of the proposed precoding algorithm is that the advantage of the legitimate link to Bob on the strong modes is exploited while the weak modes are filled with AN. Therefore, we order the channel strengths of the modes to Bob in decreasing order, i.e., $d_1^2 \geq d_2^2 \geq ... \geq d_N^2 \geq 0$. The power allocation works as follows 
\begin{eqnarray}
    p_i = \begin{cases} c & d_i^2 > \Theta \\
    0 & \textrm{otherwise} \end{cases},
\end{eqnarray}
for a threshold $\Theta>0$ which is an optimization parameter. The constant $c$ is chosen to satisfy the power constraint.
The directions (or modes), where zero signal power is allocated, are filled with AN, i.e., 
\begin{eqnarray}
    \beta_i = \begin{cases} \gamma & p_i = 0 \\
    0 & \textrm{otherwise} \end{cases},
\end{eqnarray}
where $\gamma$ is chosen to satisfy the power constraint. The two constants $c$ and $\gamma$ correspond to the power allocated to the signal modes and AN modes, respectively. The number of signal modes is $S=\norm{\mat{p}}_0$ and the number of AN modes is $N-S=A=\norm{\mat{\beta}}_0$. In addition to the threshold $\Theta$, another optimization variable is the power split $0\leq\tau\leq 1$ between signal and AN with%
\begin{eqnarray}
    c = \frac{\tau P}{S}, \qquad \gamma = \frac{(1-\tau)P}{A} \label{eq:popt}.
\end{eqnarray}

\begin{figure}[htbp!] 
	\pgfplotsset{plot coordinates/math parser=false}
	\centering
	\includegraphics[scale=1]{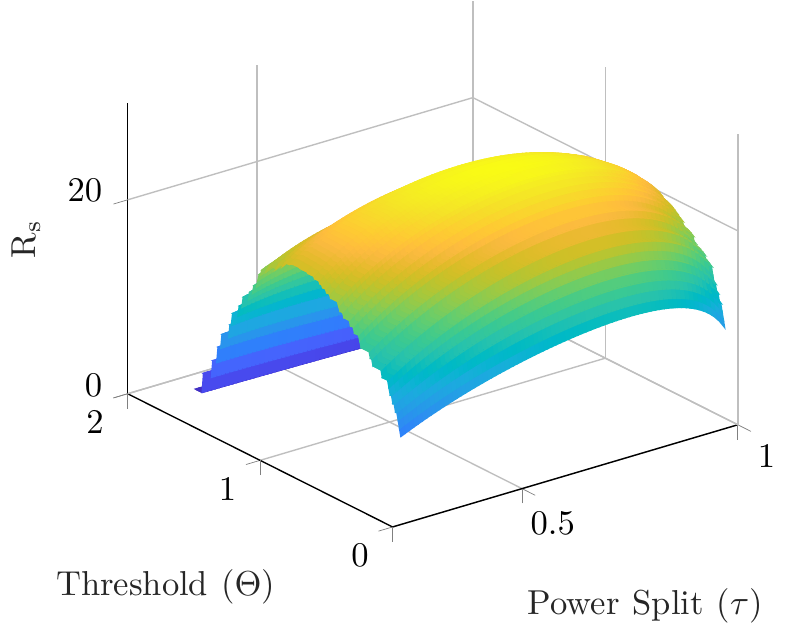}
	\caption{Illustration of unimodality/monotonicity at mid (5dB) SNR.}
	\label{fig:unimodal_mid}
\end{figure}

The remaining programming problem is to optimize the secrecy rate over the threshold $\Theta$ and over the power allocation $\tau$. Our proposed algorithm works on the assumption that the optimal power split between the information bearing signal and artificial noise power and the optimal number of signal modes are unique and the secrecy rate from \eqref{eq:objSecRate} is unimodal.

This assumption is illustrated by the numerical simulation shown in Figure~\ref{fig:unimodal_mid} for medium SNR. Our numerical evidence suggests that the unimodality property holds true for small and high SNR, too. In \cite{Lonnstrom2021}, we provide analytical proofs for this assumption. As a result, the algorithm proposed in Algorithm~\ref{alg:SVD_threshold} finds the optimal threshold $\Theta^*$ and power split $\tau^*$.

\begin{algorithm}
	\caption{Greedy AN Allocation}\label{alg:SVD_threshold}
	\begin{algorithmic}[1]
		\Require $SNR, MDL_{dB}, \mathbf{H}$
		\Procedure{Power Alloc. Calc}{$SNR, MDL_{dB}, \mathbf{H}$}
		\State Decompose (SVD) Bob's channel $\mathbf{W\Sigma V}^\dag=\mathbf{H}$;
		\State Set Threshold ($\Theta$) s.t. only strongest mode (based on eigenvalues) is higher
		\State Compute $\mathbf{F_s} = \mathbf{V} \diag({p_1,...,p_N}) \mathbf{W}^\dag $ %
		\State $\mathbf{Q_s} = P_sN (\mathbf{F_s}\mathbf{F_s}^\dag)/ \norm{\mathbf{F_s}\mathbf{F_s}^\dag}^2$;
		\State $\mathbf{Q_a} = 0$;
		\For{{each channel realization of Eve}}
		\State Create MDL matrix $\mathbf{L}$;
		\State Create Eve's Channel $\mathbf{G = \sqrt{\mathbf{L}}HU}$;
		\State Calculate $R_s$ according to (\ref{eq:capacityBob}) and (\ref{eq:capacityEve});
		\EndFor
		\State Save mean $R_s$  and corresponding $\mathbf{Q_s}$ and $\mathbf{Q_a}$
		\State Increment AN power (decrease $\tau$)
		\State Precoding $\mathbf{F_s} = \mathbf{V} \diag({p_1,...,p_N}) \mathbf{W}^\dag$ and $\mathbf{F_a} = \mathbf{V} \diag{(\beta_1,...,\beta_N)} \mathbf{W}^\dag $ 
		\State Calc $\mathbf{Q_s} = P_sN (\mathbf{F_s}\mathbf{F_s}^\dag)/\norm{\mathbf{F_s}\mathbf{F_s}^\dag}^2$;
		\State Calc $\mathbf{Q_a} = P_aN (\mathbf{F_a}\mathbf{F_a}^\dag)/\norm{\mathbf{F_a}\mathbf{F_a}^\dag}^2$;
		\State Repeat 7-16
		\State If mean $R_s$ larger than $R_s$ from previous, save $R_s$ and corresponding $\mathbf{Q_s}$, $\mathbf{Q_a}$ and repeat 13-17 until mean $R_s$ does not improve
		\State Decrease $\Theta$ and repeat 4-18 until $R_s$ no longer increases 
        \State \Return $R_s$, $\mathbf{Q_s}$, $\mathbf{Q_a}$
		\EndProcedure
	\end{algorithmic}
\end{algorithm}

\section{Numerical Illustrations}

First we show a measured MMF channel matrix. Next, we perform numerical simulations to show the achievable average secrecy rate. A performance gain of the proposed methods compared to the baseline scheme is shown. Furthermore, we compare with lower and upper bounds. 

\subsection{Measured Channels}
\begin{figure}[htbp!] 
	\pgfplotsset{plot coordinates/math parser=false}
	\centering
	\begin{tikzpicture}

\begin{axis}[%
width=0.7\linewidth,
scale only axis,
point meta min=0.00165256219920567,
point meta max=0.479691503806431,
axis on top,
xmin=0.5,
xmax=55.5,
xlabel={Mode Out},
y dir=reverse,
ymin=0.5,
ymax=55.5,
ylabel={Mode In},
axis background/.style={fill=white},
legend style={legend cell align=left, align=left, draw=white!15!black},
colormap/viridis,
colorbar,
colorbar style={
	title=Color key,
	ticklabel style={font=\footnotesize},
	ylabel=Channel Gain,
	font=\footnotesize,
	ytick={0,0.1,...,.5},
	yticklabel style={
		text width=1.0em,
		align=right,
	},
colorbar/width=2.5mm
}
]
\addplot [forget plot] graphics [xmin=0.5, xmax=55.5, ymin=0.5, ymax=55.5] {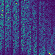};
\end{axis}

\end{tikzpicture}%
	\caption{Channel gains of the measured channel from Alice to Bob $\mat{H}$.} %
	\label{fig:measuredChannel}
\end{figure}

Using adaptive optics in the form of a spatial light modulator (SLM), the individual modes of MMFs were sequentially excited. The scattered light patterns at the fiber output were recorded holographically with a camera and then decomposed into the individual mode components using a mode decomposition algorithm \cite{Rothe2019a}. The complex mode components are entries of the transmission matrix. An example of the measured channel gains for a 55 mode fiber can be seen in Figure~\ref{fig:measuredChannel}. 

\subsection{Simulation Results}

In order to evaluate the performance of our AN noise algorithm, Monte Carlo simulations were run for various scenarios over a range of SNR. The measured channels were based on a MMF with 55 spatial modes and, unless otherwise noted, a $\text{MDL}_{dB}$ of \SI{20}{\decibel} was used for the simulations (this MDL is similar to values which were used in other works, e.g. \cite{guan2015}).

\begin{figure}[htbp!] 
	\pgfplotsset{plot coordinates/math parser=false}
	\centering
	\begin{tikzpicture}

\begin{axis}[%
width=0.7\linewidth,
scale only axis,
xmin=-15,
xmax=20,
xlabel style={font=\color{white!15!black}},
xlabel={SNR (dB)},
ymin=0,
ymax=120,
ylabel style={font=\color{white!15!black}},
ylabel={$\text{R}_\text{s}$},
axis background/.style={fill=white},
title style={font=\bfseries, align=center},
axis x line*=bottom,
axis y line*=left,
xmajorgrids,
ymajorgrids,
legend style={nodes={scale=0.65, transform shape},at={(0.058,0.584)}, anchor=south west, legend cell align=left, align=left, draw=white!15!black}
]
\addplot [color=green, line width=1.0pt]
  table[row sep=crcr]{%
-15	1.95782909751047\\
-14	2.32181639858107\\
-13	2.73851944188861\\
-12	3.21489978519131\\
-11	3.75658691825765\\
-10	4.36977050428157\\
-9	5.05439830931386\\
-8	5.81076601652141\\
-7	6.63791309313843\\
-6	7.5354783920828\\
-5	8.50293749287335\\
-4	9.53640276373371\\
-3	10.6220130058998\\
-2	11.7416753696955\\
-1	12.8712247603469\\
0	13.9953385786559\\
1	15.121526885422\\
2	16.2142259287579\\
3	17.2879244013528\\
4	18.331715739242\\
5	19.3446226882946\\
6	20.2863866873991\\
7	21.221539098216\\
8	22.1310341896758\\
9	22.9602305276096\\
10	23.8002298381945\\
11	24.627757517806\\
12	25.3917675863104\\
13	26.1010883205388\\
14	26.5850861884264\\
15	26.8699150936034\\
16	26.926291836595\\
17	26.9168981332221\\
18	26.8607581533908\\
19	26.6880047635062\\
20	26.4050765974778\\
};
\addlegendentry{Waterfilling (mean)}

\addplot [color=green, dashed, line width=1.0pt]
  table[row sep=crcr]{%
-15	1.71705452958974\\
-14	2.04773389816262\\
-13	2.42809966414926\\
-12	2.86863066379415\\
-11	3.35906126933187\\
-10	3.92019268713615\\
-9	4.54757700525056\\
-8	5.2450974782296\\
-7	5.98532688322789\\
-6	6.79163766667776\\
-5	7.66237445501989\\
-4	8.59385635835429\\
-3	9.57435149320056\\
-2	10.5649882637601\\
-1	11.5651192187142\\
0	12.5617446044489\\
1	13.5769320686242\\
2	14.6121296209376\\
3	15.6766736851219\\
4	16.7356530726955\\
5	17.5930764628721\\
6	18.4666819468137\\
7	19.3860886288534\\
8	20.2821356350014\\
9	21.0698678836988\\
10	21.8224336829601\\
11	22.5260304485014\\
12	23.2559081770497\\
13	23.9230768038712\\
14	24.2135042226234\\
15	24.3529066587879\\
16	24.2347442096602\\
17	24.104725031338\\
18	24.085923228032\\
19	24.0414091594444\\
20	23.8746593350942\\
};
\addlegendentry{Waterfilling (worst-case)}

\addplot [color=blue, dashed, line width=1.0pt]
  table[row sep=crcr]{%
-15	1.70606022371035\\
-14	2.03442414531672\\
-13	2.40742910929545\\
-12	2.82885586002237\\
-11	3.32882090622713\\
-10	3.88138184784714\\
-9	4.51241160136238\\
-8	5.22125987208901\\
-7	5.9710687956086\\
-6	6.79126695258428\\
-5	7.66825741638471\\
-4	8.60079820181014\\
-3	9.74213626805997\\
-2	10.9177163616753\\
-1	12.3374381770996\\
0	13.9308833754383\\
1	15.6789681849781\\
2	17.5889509912265\\
3	19.7236587725253\\
4	22.0400114941398\\
5	24.5730790788789\\
6	27.3222148337264\\
7	29.9955465237045\\
8	33.0619913010524\\
9	36.2130069959488\\
10	39.4270039154943\\
11	42.7095119870754\\
12	45.9978192999227\\
13	49.2392617851363\\
14	52.4273961915822\\
15	55.5570692343508\\
16	58.6178169569905\\
17	61.6053742625887\\
18	64.4838824983604\\
19	67.2467505644548\\
20	69.8848671231798\\
};
\addlegendentry{Greedy AN (worst-case)}

\addplot [color=blue, line width=1.0pt]
  table[row sep=crcr]{%
-15	1.93502663488057\\
-14	2.29016299486512\\
-13	2.70252593840029\\
-12	3.16896905930955\\
-11	3.69144326761117\\
-10	4.28021368003485\\
-9	4.94659469846921\\
-8	5.69590218315034\\
-7	6.51110334501468\\
-6	7.40365766293712\\
-5	8.33588299739037\\
-4	9.3509090168224\\
-3	10.4255718570105\\
-2	11.6365830766066\\
-1	13.0617442857763\\
0	14.6671564777206\\
1	16.4838980277292\\
2	18.4814513475367\\
3	20.6869167159065\\
4	23.0802591378034\\
5	25.6491793562329\\
6	28.4286726337459\\
7	31.3574287550716\\
8	34.4363329454068\\
9	37.6407002420906\\
10	40.8976649752015\\
11	44.1799274690853\\
12	47.4627509834848\\
13	50.761776449686\\
14	54.0169870328977\\
15	57.2246911056371\\
16	60.3560011562699\\
17	63.3863995834096\\
18	66.3017238064414\\
19	69.0998631192473\\
20	71.7756164253361\\
};
\addlegendentry{Greedy AN (mean)}

\addplot [color=red, line width=1.0pt]
table[row sep=crcr]{%
-15	3.10584329710462\\
-14	3.75104874271848\\
-13	4.50957443728034\\
-12	5.38088408351564\\
-11	6.40503691844155\\
-10	7.66556250246942\\
-9	9.10285098893173\\
-8	10.7518139499717\\
-7	12.6353339314269\\
-6	14.7213543250679\\
-5	17.1307051854138\\
-4	19.8301477371409\\
-3	22.7540110785825\\
-2	25.9228065275721\\
-1	29.4192907035054\\
0	33.1011356138706\\
1	36.9346192651337\\
2	40.8840371004089\\
3	44.9131405672345\\
4	49.0605860087407\\
5	53.2646731349162\\
6	57.4490606308866\\
7	61.5828469711154\\
8	65.6390354204845\\
9	69.5946808173409\\
10	73.430836207261\\
11	77.1323565192964\\
12	80.6876113392479\\
13	84.0881489372056\\
14	87.422320902439\\
15	90.5989860484267\\
16	93.6077518695463\\
17	96.4491703288622\\
18	99.1253424476202\\
19	101.639627810229\\
20	103.996384603367\\
};
\addlegendentry{Upper Bound}

\addplot [color=red, dashed, line width=1.0pt]
table[row sep=crcr]{%
-15	0\\
-14	0\\
-13	0\\
-12	0\\
-11	0\\
-10	0\\
-9	0.0143702519354107\\
-8	0.0492730600719877\\
-7	0.100604710466181\\
-6	0.172250865380402\\
-5	0.267880684538742\\
-4	0.390525697576116\\
-3	0.542158619405406\\
-2	0.723341166624092\\
-1	0.933000582460807\\
0	1.16837207215606\\
1	1.42511677330211\\
2	1.69760004027107\\
3	1.97929657620799\\
4	2.26327690049143\\
5	2.54272183062135\\
6	2.81140772506188\\
7	3.06410734087711\\
8	3.29686209605494\\
9	3.50710168913487\\
10	3.693612738963\\
11	3.85638261086481\\
12	3.99636113031437\\
13	4.11518781794315\\
14	4.21492637377374\\
15	4.29783547231858\\
16	4.36619054979204\\
17	4.42215907457648\\
18	4.46772366643802\\
19	4.50464342703876\\
20	4.53444299555676\\
};
\addlegendentry{Lower Bound}

\addplot [color=black, line width=1.0pt]
table[row sep=crcr]{%
	-15.0000268949424	1.957121566449206\\
	-11.0000268949424	3.695356700230502\\
	-7.00002689494244	6.283279417032571\\
	-5.00002689494244	7.788951924482973\\
	-3.00002689494244	9.309011143146943\\
	-1.00002689494244	10.726643666446780\\
	0.999973105057563	11.944775874605273\\
	8.99997310505756	14.462567807440445\\
	18.9999731050576	15.028730446542456\\
};
\addlegendentry{Ergodic Secrecy Rate \eqref{eq:jenSecRate}}

\end{axis}

\end{tikzpicture}%
	\caption{Comparison of different precoding algorithms and their effect on the secrecy rate using Monte Carlo simulations.}
	\label{fig:precodeCompMC}
\end{figure}
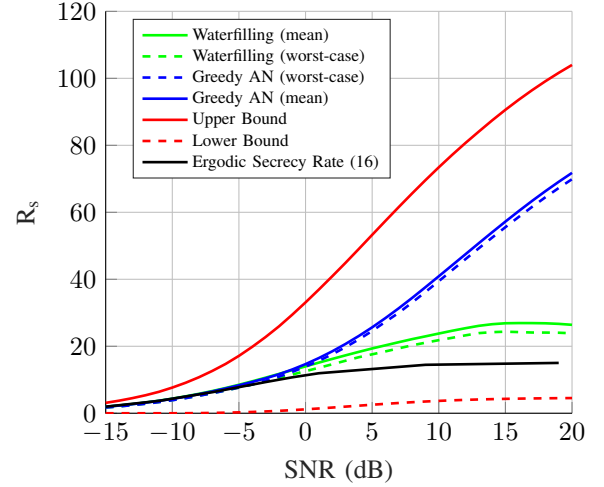

The solid and dashed blue curve in Figure~\ref{fig:precodeCompMC} shows the performance of the proposed Greedy AN algorithm over a range of SNR.
The dashed blue line represents the worst-case secrecy rate which was seen over 20k channel realizations.
As a comparison, we calculated the secrecy rates for a peaceful water filling algorithm \cite{Tse2005} which maximizes the rate between Alice and Bob without taking the eavesdropper's channel into account as seen in the solid and dashed green curves.
Of course the water-filling algorithm is not optimal for power allocation for the purposes of physical layer security, nonetheless it is used here to demonstrate the gain which can be achieved using our AN algorithm.
At around \SI{0}{\decibel} SNR, the greedy AN algorithm shows a visible gain in achievable secrecy rate.
By \SIlist{10;15}{\decibel} SNR the greedy AN algorithm outperforms a peaceful water-filling algorithm by over a factor 1.5x and 2x, respectively. 
The greedy AN algorithm also performs better than the the lower bound calculated for the pessimistic ergodic secrecy rate according to \eqref{eq:jenSecRate} which was based on Jensen's inequality. 

The upper and lower bounds, as seen by the solid and dashed red lines respectively show a relatively wide range for the possible secrecy rates, however, the variance of Greedy AN algorithm after 20k random channel realizations shows that this lower bound is quite unlikely in a real-world situation. Since the water-filling algorithm only concentrates on maximizing the rate between Alice and Bob without taking the presence of an eavesdropper into account it is obviously not an optimal solution for optimizing secrecy.  

\section{Conclusions}

The development of precoding and power allocation to wiretap-coded transmission over a MMF channel is developed for imperfect CSI at the transmitter and AN. It is demonstrated based on measured channel matrices that we can achieve a positive average secrecy rate even in channels where the advantage of the legitimate channel is realized only over a small subset of modes. Compared to the state of the art SVD precoding, the AN-based algorithm scales well with the SNR with growing gains. Furthermore, the computed lower bound for the achievable secrecy rate shows that even under the worst case wiretap channel matrix, positive secrecy rates can be supported. This motivates us to consider outage secrecy rates as well as zero-outage secrecy rates in our future work. 

\section*{Acknowledgements}
The work of A. Lonnstrom and E. Jorswieck is supported in part by the German Research Foundation (DFG) under grant JO~801/25-1. The work of Stefan Rothe and Jürgen Czarske is supported in part by DFG under grant CZ 55/42-1.

\printbibliography

\end{document}